# A coronary artery phantom for task-based CT performance assessment and a comparative study of clinical CT, photon counting CT, and micro CT

[1]Jed D. Pack, [1]Paul Fitzgerald, [1]Stephen Araujo, [1]Ying Fan, [2]Grant Stevens, [3]Jonathan Gerdes, [4]Adam Wang, [4]Koen Nieman, [5]Ge Wang, [1]Bruno De Man

[1]GE HealthCare—Technology & Innovation Center, Niskayuna, NY

[2]GE HealthCare, Menlo Park, CA

[3]Phantom Laboratories, Greenwich, NY

[4]Stanford University, Palo Alto, CA

[5]Rensselaer Polytechnic Institute, Troy, New York

Abstract

While drastic improvements in CT technology have occurred in the past 25 years, spatial resolution is one area where progress has been limited until recently.  New photon counting CT systems, are capable of much better spatial resolution than their (energy integrating) predecessors.  These improvements have the potential to improve the evaluation obstructive coronary artery disease by enabling more accurate delineation between calcified plaque and coronary vessel lumen.  A new set of vessel phantoms has been designed and manufactured for quantifying this improvement.  Comparisons are made between an existing clinical CT system, a prototype photon counting system, with images from a micro CT system being used as the gold standard.  Scans were made of the same objects on all three systems.  The resulting images were registered and the luminal cross section areas were compared.  Luminal cross-sections near calcified plaques were reduced due to blooming, but this effect was much less pronounced in images from the prototype photon counting system as compared to the images from the clinical CT system.

1. Introduction

Since the advent of multi-detector row CT (MDCT) systems 25 years ago, clinical CT systems have made remarkable progress in terms of speed, volume coverage, temporal resolution, reliability, dose efficiency, workflow, and spectral capability.  At the same time, the spatial resolution of most clinical CT scanners today is not dramatically better than that of the first MDCT systems from 1998, with a few exceptions.  Recently, clinical photon counting CT (PCCT) systems have shown promise for markedly better spatial resolution capability.  This paper introduces a set of new phantoms designed to test the impact of spatial resolution on coronary CT angiography (CCTA) in a clinically relevant task-driven way.

To motivate the phantom selection, it is important to give some background on the clinical need for ultra-high spatial resolution. Many clinical tasks today do not require ultra-high spatial resolution.  In fact, many clinical images produced today have a slice thickness of at least 2 mm.  While this choice degrades spatial resolution relative to even the current system capability, it helps to manage the number of images that a radiologist must review for this large set of clinical tasks for which the highest spatial resolution adds little, if any, value.

Regardless of this, there are still some very important clinical tasks for which ultra-high spatial resolution is critical. For example, evidence of the progression of coronary artery disease (CAD) can be seen in the form of calcified and non-calcified plaques built up in the coronary arteries. These plaques can obstruct blood flow to the heart muscle, leading to chest pain as well as adverse events (e.g., a heart attack). Determining whether such obstructions are hemodynamically significant is a key to guiding treatment decisions (e.g., catheterization and stent placement). However, such evaluation is challenging due to the limitations of system spatial resolution today in the form of calcium blooming [1] and blurring-induced reconstructed contrast reduction in narrow-luminal coronary sections [2]. Even with the spatial resolution of current scanners, CCTA is considered a key first line test for evaluating CAD [3]. Furthermore, use of CCTA images for fluid dynamics simulations in order to predict the fractional-flow reserve are becoming more common and rely heavily on an accurate geometrical model of the coronaries, which can be hampered by inadequate spatial resolution.

Given the importance of spatial resolution in CCTA, we have chosen to design a set of new anatomically realistic coronary artery phantoms specifically to test image quality in the context of obstructive CAD. To fully separate the effect of motion artifacts (which are beyond the scope of this work) from the effect of the inherent system spatial resolution, we scan our phantoms in static mode (without any motion). Optimization of motion artifact reduction algorithms for ultrahigh resolution systems will also be required to fully realize the advantages of improved spatial resolution in clinical practice.

The paper is organized as follows. In section 2, we describe our phantom design, detail the phantom production, and summarize our protocols for scanning the phantom on three devices: a Revolution APEX™ clinical CT scanner, a BHGE Phoenix Vtomex M micro-CT scanner, and the new Deep Silicon PCCT system located at Stanford University. In section 3, we show results, including several image comparisons, and a first-pass quantitative analysis of the resolution characteristics of the three systems based on the new phantom. Finally, in section 4, we highlight a few interesting points of discussion and concluding remarks.

2. Methods
2.1. Phantom Design

The original basis for our phantom designs is a set of clinical CCTA scans. The longest continuous coronary artery was automatically selected from each exam and an automated centerline extraction and lumen segmentation was performed. A three-dimensional (3D) mesh of each of these coronaries was displayed and twenty sections were manually selected. Rationale for selecting the twenty chosen sections included a desire for a variety of luminal diameters from 2 to 5 mm as well as interesting, but not extreme curvature. The centerlines were represented by a smooth sampling of points spaced by 0.2mm. A local two-dimensional (u,v) coordinate system was defined for each centerline point in the plane orthogonal to the tangent direction. The outer boundary of the intersection of the lumen segmentation with this plane was defined as a curve using fifty (u,v) coordinate pairs. These sets of 50 points (one set for each centerline point) comprised the 3D lumen mesh.

To conform to the requirements of the manufacturing process, the centerlines needed to be constrained to a single plane. The process by which this was done will now be described in detail. Mathematically, constraining the centerlines to a plane means that the torsion of the centerline curves needed to be set to zero everywhere. For each set of three adjacent points on a curve, the curvature is related to the angle between the directions of the two corresponding line-segments. These two line-segments lie in a plane. Meanwhile, the next set of three adjacent points (constructed by removing the first of the three points and appending the point immediately after the third point) also lie in a plane. The angle between these two planes is related to the torsion of the curve. We forced the torsion to be zero by stepping down each segment of the curve (beginning at the second segment) and rotating the entire distal part of the curve about the axis of the selected segment by the smallest possible angle that ensured that the subsequent point belonged to the same plane as the prior three points. For example, given N centerline points ($p_1$, $p_2$, … $p_N$), the first three points necessarily belong to a plane, so in the first step of this process, all distal points ($p_4$, $p_5$, …, $p_N$) are rotated about the axis connecting $p_2$ to $p_3$ such that $p_4$ also lies in that same plane. This process preserves as much of the overall shape of the curve as possible while removing any torsion and constraining the curve to a single plane.

A second constraint of our design is that the processed vessel sections should fit within a cylinder of length 7.5 cm and diameter 2.5 cm. Three of the twenty sections that fit this requirement were selected for further processing.

The next step in our process was to introduce holes or pits in the surface of the vessel structure to make room for calcified plaques to be added later (Figure 1 a). We took our inspiration for this process from a system of grading calcium deposits used in [4] wherein each calcified plaque is characterized as being one of four types (I-IV) depending on the number of quadrants occupied by the plaque in the cross-sectional view. We introduced pits in seven or eight regions for each of our three vessels. These pits wrapped around the outer edge of our vessel to cover either one, two, three or four quadrants, with the last type (type IV) wrapping completely around the outer edge of the lumen. We included two subtypes for type II, wherein type IIa plaques covered two consecutive quadrants and type IIb plaques being comprised of two small opposing plaques residing in opposing quadrants (Figure 1b). After introducing these pits, the vessel meshes were exported as stl files and provided to a manufacturer (Phantom Laboratory, Greenwich NY) for production.

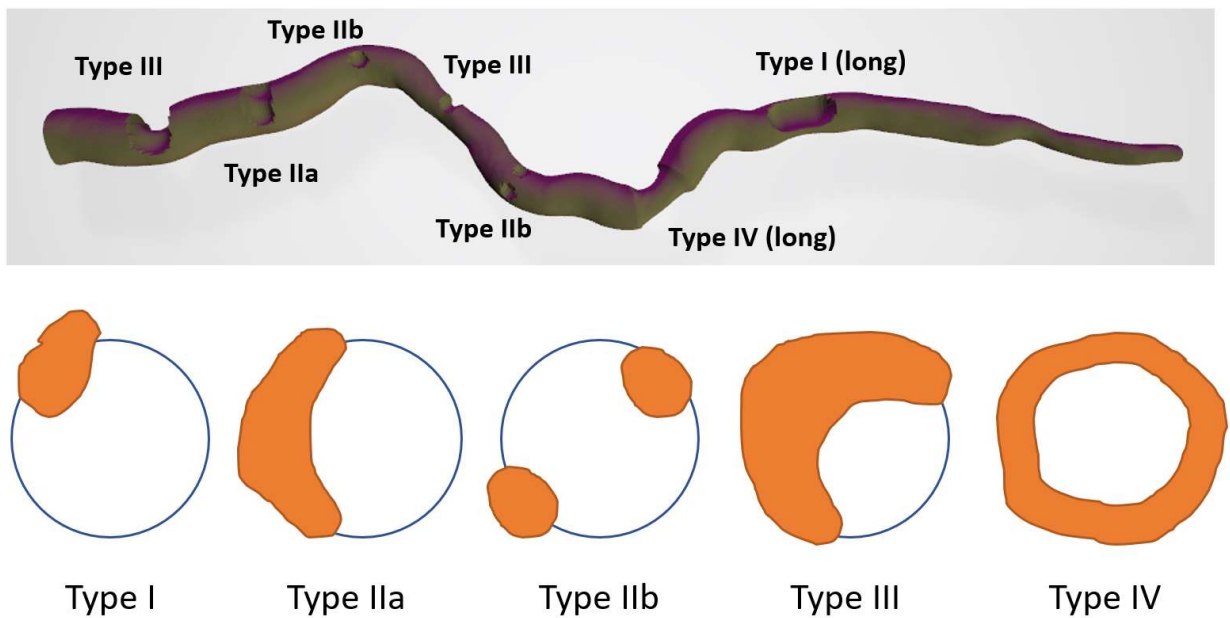

Figure 1: Example of a 3D coronary artery lumen model. (a) Pits around the outer edge of the vessel make room for calcifications. (b) Various types of calcifications are defined based on the number of quadrants they occupy.

2.2 Phantom Production

The three stl files produced in section 2.1 were printed using a high-resolution 3D printer to produce 3 vessel-shaped objects in the shape of the digital phantoms. Next, a rubber mold was fashioned around each of the three vessels and each mold was divided in half in the plane of the final vessel centerline. The original 3D printed material was then removed from the mold, leaving a (negative) void in the shape of the original stl file between the two halves of the mold.  After reassembling the two halves, a casting material was produced; an iodinated contrast agent was thoroughly mixed with liquefied plastic to increase its linear x-ray attenuation to mimic iodinated blood.  This material was introduced to the mold to produce (after curing) a (positive) vessel shape. This casting process was repeated twice for each of the three molds, resulting in a total of six vessels made from three concentrations of iodine contrast.  The iodine mixtures contained 10, 15 and 20 mg/ml of iodine.

Once the six vessel shapes were in hand, the next step was to manually apply a calcium/epoxy paste mixture to fill the voids that were created to host calcified material.  The fraction of calcium salts in the mixture was chosen to produce an x-ray attenuation value that was representative of clinical calcified plaques (~1,100 HU).  Next, each of the six vessels was placed inside a cylindrical mold of diameter 2.5 cm and length 7.5 cm.  Liquified plastic (~50 HU) was then cast into these cylindrical molds around the vessels.  Once cured, the solid cylinders containing the vessels were removed from the mold.  An image of three of the phantoms is shown in Figure 2.

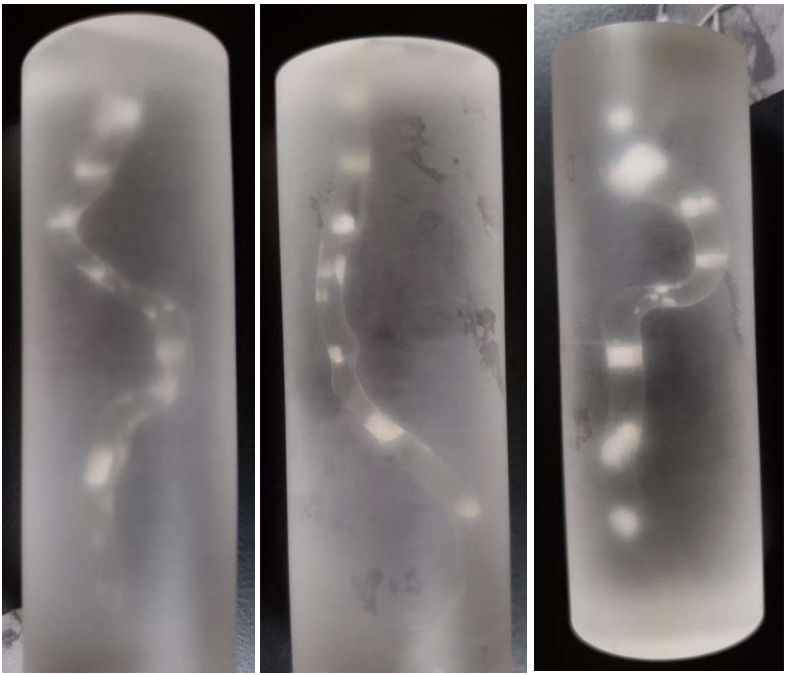

*Figure 2: Three of the six vessel phantoms are shown in these photographs.  They have reaslitic shape and iodine concentration, have 7 to 8 calcifications each, and are cast in a plastic cylinder.*

2.2. Scanning

2.2.1. Energy Integrating Clinical CT

The phantoms were scanned two at a time on a Revolution Apex scanner inside a 20 cm deep QRM cardiothoracic phantom.  We used an axial scan protocol with a 16cm z-coverage (256 x 0.625mm), 100kVp, 470 mA, and a 1.0 s gantry period.  The QRM phantom has a 10cm diameter cylindrical cavity.  We fixed two 2.5cm diameter phantoms inside a 10cm diameter water phantom using a plastic frame to separate them in the center of the phantom (see Figure 3).

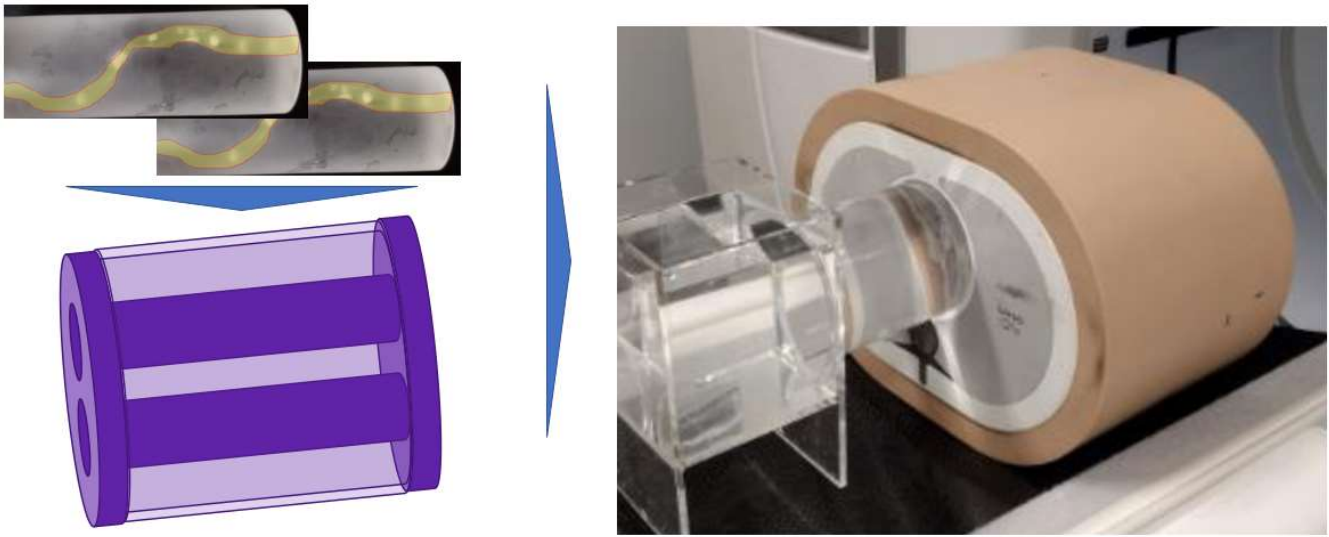

*Figure 3: Two 2.5 cm diameter vessel phantoms are fixed inside a 20 cm diameter water phantom, which in turn is inserted in a QRM phantom.*

### 2.2.2. Deep-Silicon Photon-Counting CT

The phantoms were also scanned on a new deep-silicon PCCT scanner prototype located at Stanford University. The cylinders were inserted into 3.5cm diameter holes in a gammex phantom for scanning. The scan was done in axial mode with 120 kVp, 280 mA, 1 s gantry rotation, extra small focal spot, and 40 mm z collimation. Two axial scans were performed to cover the entire z-extent of the cylinders in two slabs. The axial images were reconstructed using FBP with a 1024x1024 matrix covering a 20 cm field-of-view.

### 2.2.3. Micro CT

Finally, the bare phantom cylinders were scanned on a BHGE Phoenix Vtomex M micro-CT system in order to provide extremely high-resolution images of the calcification and lumen geometries, serving as ground-truth images. The scans were done in an axial mode with two scans needed to cover the full 7.5 cm length of the phantoms. The system has two X-ray tubes and the nano-focus X-ray tube was used here. The scan settings were 120 kVp, 160 µA, 333 ms dwell time (repeated 7 times and averaged), 1000 views, and a 45-minute scan time (for each of the two axial scans per phantom).

Scan and reconstruction parameters for all three systems are listed in table I.

|  | kVp | mA | Scan time (s) | Focal spot | Axial Coverage | Voxel size (mm) | Reconstruction |
| --- | --- | --- | --- | --- | --- | --- | --- |
| Clinical CT | 100 | 420 | 1.0 | S | 160 mm | 0.4x0.4x0.625 | ASiR-V 50 |
| Deep Silicon PCCT | 120 | 280 | 1.0 (x2) | XS | 40 mm (x2) | 0.2x0.2x0.42 | FBP |
| Micro CT | 120 | 0.333 | 2700 (x2) | Nano | 37 mm (x2) | 0.02x0.02x0.02 | FBP |

*Table 1: Scan and reconstruction parameters for phantom scans with three CT scanner types.*

3. Results

3.1. Images

The resulting images from all three systems are shown in figure 4. The images are separated into a left hand group and a right hand group. Images on the left group are shown with a wide HU window width (1200 HU), while images on the right are identical to the images on the left except that they are displayed with a narrower window width (800 HU), which is a typical window setting for reviewing CCTA images. Within each group, the three columns contain images from the three scanners as indicated (Apex, Deep Silicon, and microCT, respectively). Meanwhile, the four rows highlight four corresponding locations (L1-L4) of calcified material (representing calcified plaques of different types).

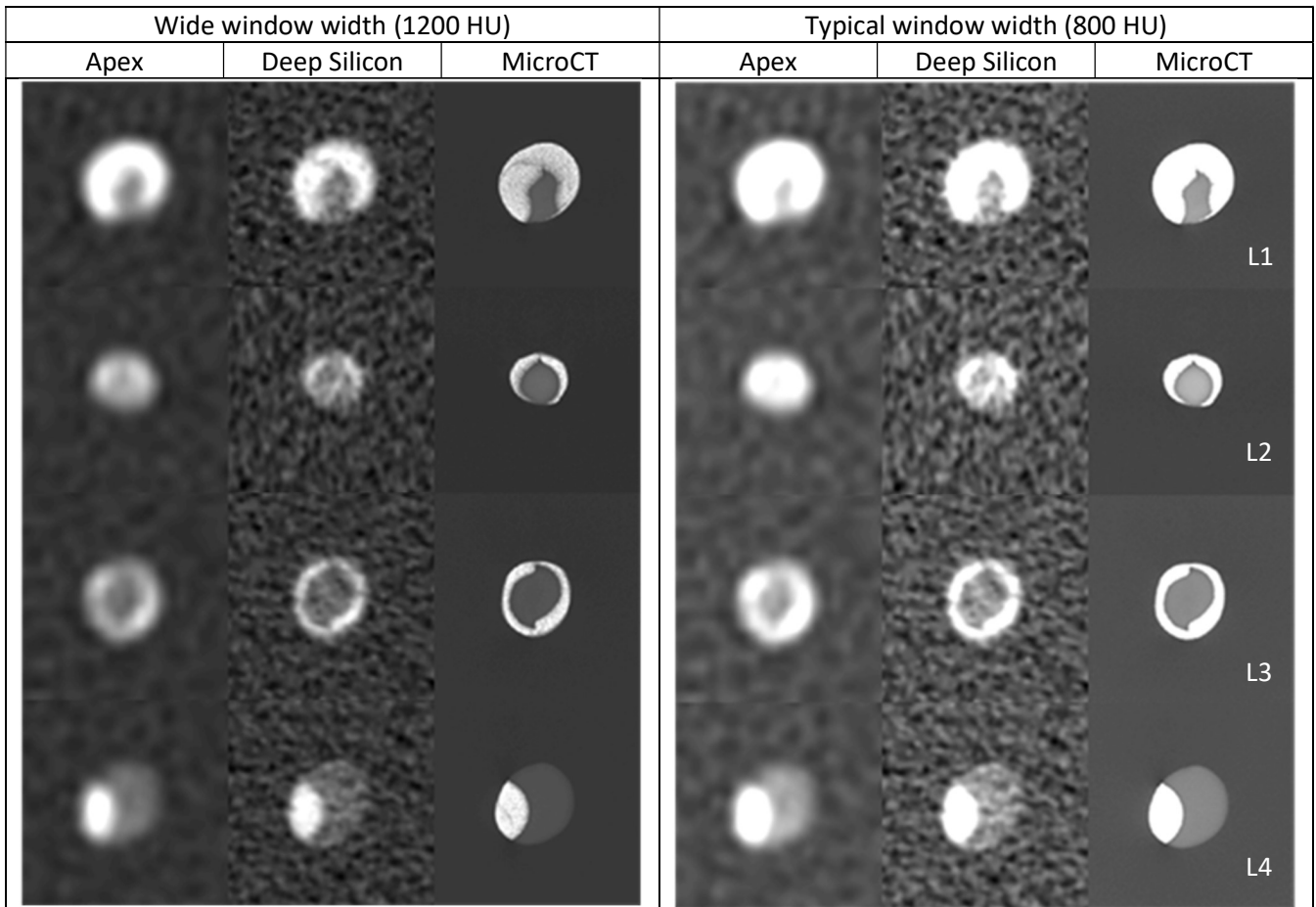

*Figure 4: Example CT images of the vessel phantoms for clinical CT, deep-silicon photon-counting CT, and micro-CT. Image volumes were manually registered to show approximately the same cross-sections.*

3.2. Image Analysis Results

As a first step toward quantifying the calcium blooming effect in our images, the gold-standard microCT images were segmented manually to delineate the lumen region and the calcified region. These regions are shown by red lines in figure 5.

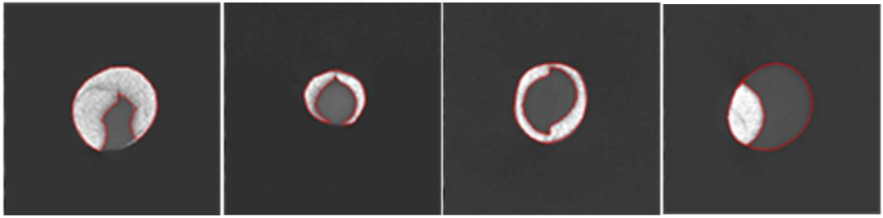

Figure 5: Manually delineation of the lumen and calcified plaque regions are shown by red lines.

A simple threshold was then applied to the images from the other two scanners (Apex and Deep Silicon) based on the idea that voxels above this threshold might be perceived as being part of the calcified plaque by a human observer, while voxels below this threshold would be perceived as being part of the lumen. Comparisons were then made between the lumen cross-sectional area (CSA) determined using this threshold and the "true" lumen CSA. The ratio between the CSA measured in this way and the true CSA was defined as the relative CSA (rCSA).

$$rCSA = \frac{CSA}{CSA_{True}}$$

Only voxels within the original (ground truth) lumen mask were analyzed in this way. As a result, all rCSA values are less than or equal one, with a value near one indicating very little calcium blooming due to a high scanner spatial resolution. The rCSA depends on the shape of the calcium and the lumen. For example, a type III or IV calcification may cause much more blooming than a type I calcification, resulting in a smaller rCSA value. In addition, the rCSA also depends on the selected threshold value. In a similar way, the perceived size of a calcified plaque depends on the window width setting used to view an image volume.

The relative cross-sectional area (rCSA) values computed using our two different thresholds (500 HU and 640 HU) are given in the table below:

|  | L1 | L2 | L3 | L4 |
| --- | --- | --- | --- | --- |
| rCSA (DS/Apex) Thresh=500HU | 0.913/0.439 | 0.727/0.036 | 0.988/0.689 | 0.986/0.951 |
| rCSA (DS/Apex) Thresh=640HU | 0.999/0.823 | 0.976/0.595 | 1.000/0.970 | 0.999/0.999 |

Table 1: The relative cross-sectional area (rCSA) values are given for the four different image locations (L1-L4) shown in figures 4 and 5. The rCSA values were computed using two different thresholds:

4. Discussions and conclusions

A visual inspection of the images in figure 4 shows a reduction in calcium blooming for the Deep Silicon photon counting scanner relative to the clinical CT scanner, indicating a higher spatial resolution (as expected). In addition, it is clear by comparing the images at the two window width settings that the apparent blooming is reduced by a substantial amount in the wider window width as compared to the typical window with setting. While use of a wide window setting can help for visual sizing of calcified plaques, a wide window width is not ideal for other tasks. Quantification of the cross-sectional area using a simple threshold-based approach also demonstrated that the Deep Silicon photon counting scanner gave more accurate values for the luminal cross-section. In conclusion, the higher spatial resolution of the prototype Deep Silicon photon counting scanner has been shown to improve luminal sizing near calcified plaques by reducing the impact of the calcium "blooming" effect.